%% file: Template.tex
\documentclass{article}
\usepackage{spconf,amsmath,graphicx,hyperref}

\apptocmd{\thebibliography}{\setlength{\itemsep}{0.25ex}\setlength{\parskip}{0pt}}{}{}

\usepackage{cite}
\usepackage{amsmath,amssymb,amsfonts}
\usepackage{algorithmic}
\usepackage{graphicx}
\usepackage{booktabs,siunitx}
\usepackage{textcomp}
\usepackage{multirow}
\usepackage{multicol}
\usepackage{xcolor}
\usepackage{subcaption}
\usepackage{comment}
\usepackage{enumitem}

\usepackage{pgfplots}
\usepackage{pgfplotstable}

\usepgfplotslibrary{colormaps}
\pgfplotsset{colormap/jet,}

\usepackage{pdfpages}
 \usepackage{float}
\pgfplotsset{compat=1.18} 

\usepackage{url}
 
\usepgfplotslibrary{groupplots,external}
\usepackage{tikz}
\usetikzlibrary{bayesnet}
\usetikzlibrary{arrows, automata}
\usetikzlibrary{arrows.meta}
\usetikzlibrary{shapes, arrows, positioning, calc}
\usetikzlibrary{shadows.blur}
\usetikzlibrary{shapes.symbols}

\pgfplotsset{compat=newest}%
\usetikzlibrary{pgfplots.groupplots}
\usetikzlibrary{pgfplots.colormaps}%
 \usetikzlibrary{arrows.meta} 
 
\usepgfplotslibrary{groupplots}

\def\BibTeX{{\rm B\kern-.05em{\sc i\kern-.025em b}\kern-.08em
    T\kern-.1667em\lower.7ex\hbox{E}\kern-.125emX}}

\usepackage{pgfplots}
\usepgfplotslibrary{groupplots,colormaps}
\pgfplotsset{compat=1.18}

\pgfplotsset{
    colormap={myrevjet}{
      indices of colormap={\pgfplotscolormaplastindexof{jet},...,0 of jet}
    }
}
\pgfplotsset{compat=1.18}
\usetikzlibrary{matrix}
\usetikzlibrary{positioning,calc}
\usepgfplotslibrary{colormaps}
\usepackage{tikz} 

\pgfplotsset{compat=1.18}

\usepackage{makecell} 
\usepackage{array} 

\usepackage{threeparttable}

\newcolumntype{C}[1]{>{\centering\arraybackslash}b{#1}} 
\newcolumntype{M}[1]{>{\centering\arraybackslash}m{#1}}

\usepackage{tabularx}

\makeatletter
\renewcommand\section{\@startsection{section}{1}{\z@}%
  {-1.ex \@plus -1ex \@minus -.2ex}%
  {0.3em} 
  {\normalfont\Large\bfseries}}
\makeatother

\title{VBx for end-to-end neural and clustering-based diarization}
\name{Petr P\'{a}lka$^{1}$, Jiangyu Han$^{1}$, Marc Delcroix$^2$, Naohiro Tawara$^2$, Luk\'{a}\v{s} Burget$^{1}$}
\address{$^1$Brno University of Technology, Speech@FIT, Czechia\hspace{1cm} $^2$NTT, Inc., Japan
}

\begin{document}
\ninept

\maketitle
\begin{abstract}
We present improvements to speaker diarization in the two-stage end-to-end neural diarization with vector clustering (EEND-VC) framework. The first stage employs a Conformer-based EEND model with WavLM features to infer frame-level speaker activity within short windows. The identities and counts of global speakers are then derived in the second stage by clustering speaker embeddings across windows. The focus of this work is to improve the second stage; we filter unreliable embeddings from short segments and reassign them after clustering. We also integrate the VBx clustering to improve robustness when the number of speakers is large and individual speaking durations are limited. Evaluation on a compound benchmark spanning multiple domains is conducted without fine-tuning the EEND model or tuning clustering parameters per dataset. Despite this, the system generalizes well and matches or exceeds recent state-of-the-art performance.

\end{abstract}
\begin{keywords}
speaker diarization, EEND-VC, VBx, pyannote
\end{keywords}
\vspace{0.2em}
\section{Introduction}
\label{sec:introduction}


Speaker diarization is gradually shifting from traditional clustering-based cascaded systems~\cite{park2019auto, landini2022bayesian} toward end-to-end neural diarization (EEND) ~\cite{fujita19_interspeech,fujita2019_eend_SA, kinoshita2021integrating, kinoshita2021advances}.
End-to-end neural diarization with vector clustering (EEND-VC) \cite{kinoshita2021integrating, kinoshita2021advances} has emerged as a leading framework.
For long recordings, EEND-VC divides the input into short windows and applies EEND to each, obtaining local diarization estimates. Across-recording identities, obtained by the clustering algorithm, are then assigned to these local windows and subsequently aggregated.
This hybrid pipeline effectively handles overlapped speech and an arbitrary number of speakers, providing a practical diarization solution for real-world applications.

Pyannote \cite{bredin23_interspeech}, an open source implementation of the EEND-VC paradigm, has achieved competitive performance on many public datasets \cite{plaquet23_interspeech}. To further improve the performance of this system, researchers have introduced more advanced modules such as Mamba \cite{plaquet2024mamba} and WavLM \cite{han2025leveraging} to improve the local EEND component. However, the speaker-clustering stage has received less attention.
In pyannote, the default method is agglomerative hierarchical clustering (AHC), which could potentially be improved by adopting more advanced clustering approaches such as VBx \cite{landini2022bayesian}.

Based on a Bayesian hidden Markov model (BHMM), VBx aims to cluster speakers in a sequence of speaker embeddings (x-vectors).
It assumes that the input x-vector sequence is generated by an HMM whose speaker-specific state distributions are derived from a two-covariance probabilistic linear discriminant analysis (PLDA) model~\cite{brummer10_odyssey}.
Although VBx has shown competitive performance on many challenging benchmarks~\cite{landini2020but, landini2024diaper}, its integration into an EEND-VC framework such as pyannote is not straightforward~\cite{delcroix2023multi}.
One key reason is that the original VBx relies on an HMM to model speaker embeddings, which makes it difficult to handle embedding sequences coming from the non-contiguous segments, a situation that commonly occurs in the EEND-VC framework.

In this work, we investigate how to improve the speaker clustering of EEND-VC by incorporating VBx.
Following the pyannote pipeline, we first revisit the details of its clustering and then propose to use a simplified VBx where the original HMM is degraded into a Gaussian Mixture Model (GMM). Although simple, this idea can be effectively applied within the EEND-VC framework.
In addition, we propose filtering speaker embeddings from short segments so that subsequent clustering benefits from the remaining high-quality embeddings.
Our experiments, conducted with pyannote \cite{bredin23_interspeech, plaquet23_interspeech} and DiariZen \cite{han2025leveraging, han2025efficientgeneralizablespeakerdiarization}, are evaluated on numerous public datasets including AMI~\cite{carletta2005ami}, AISHELL-4~\cite{fu21b_interspeech}, AliMeeting~\cite{yu2022m2met}, NOTSOFAR-1~\cite{vinnikov24_interspeech}, MSDWild~\cite{liu22t_interspeech}, DIHARD3 full~\cite{ryant21_interspeech}, RAMC~\cite{yang22h_interspeech}, and VoxConverse~\cite{chung20_interspeech}.
The results show consistent improvements across all datasets.
Furthermore, when DiariZen-Large is combined with VBx, our system surpasses the most recent state-of-the-art performance on the majority of the benchmarks.

\section{Diarization Pipeline}
\label{sec:diarization_pipeline}

We follow the two-stage EEND-VC pipeline adopted in pyannote~\cite{bredin23_interspeech}. The first stage (\emph{local EEND}) produces local window-level frame-wise predictions of speaker activity, while the second stage (\emph{vector clustering}) consolidates these predictions into consistent global speaker identities across the entire recording. Our work focuses on improving the clustering stage.


The local EEND model processes the input audio in short windows or chunks. These windows are extracted with overlap, allowing the same portion of the waveform to be analyzed in multiple contexts, which leads to more reliable diarization decisions. In this work, we use the EEND model from the DiariZen~\cite{han2024leveragingselfsupervisedlearningspeaker} system, which combines a WavLM-based front end with a Conformer-based back end to predict frame-by-frame speaker activity.


In the following, we describe steps of the second stage of the EEND-VC pipeline, which are central to our approach. In our work, the clustering and reassignment of embeddings are performed using the pyannote implementation, which we take as a baseline.

\vspace{-2mm}
\subsection{Embedding extraction} 
\label{embedding_extraction}
\vspace{-1mm}

The pipeline relies on a standard speaker embedding extractor trained for single-speaker classification, in contrast to approaches that integrate embedding learning directly into diarization, where embeddings are optimized jointly with the diarization model~\cite{kinoshita2021advances}.  
Figure~\ref{fig:spk_ee_strategy} illustrates the local diarization output for a single chunk of waveform, with the maximum number of local speakers set to four. Each detected speaker is represented on a separate ``track''; for example, the first track (green) contains three active segments (1, 2, 3) not overlapping with another speaker. These segments are concatenated into a single waveform (1~\(\oplus\)~2~\(\oplus\)~3), from which a speaker embedding is extracted. A special case occurs when a speaker is active only in overlapping regions, such as the orange track in the figure. In this case, all segments are concatenated to form the input for embedding extraction. 


\subsection{Clustering algorithm}
\label{clustering_algorithm}
\vspace{-1mm}

Pyannote performs agglomerative hierarchical clustering (AHC) with centroid linkage (or unweighted pair group method using centroids (UPGMC)), merging clusters based on the cosine distance between their centroids. Each centroid is computed as the arithmetic mean of the embeddings in the cluster. This criterion often preserves small outlier clusters until later in the merging process. To mitigate this effect, a post hoc \emph{minimum--cluster--size} (\(mcs\)) threshold is applied: for each cluster~\(C\), if its size~\(|C|\), defined as the number of embeddings assigned to the cluster, is smaller than \(mcs\), the cluster is labeled as `small', and all its embeddings are reassigned to the nearest `large' cluster according to the cosine distance between the two centroids. 



\begin{figure}[!bt]
    \centering
    \input{figures/spk_ee_strategy}
    \vspace{-0.3cm}
    \caption{Colors indicate local speakers activities, while rectangles mark waveform segments concatenated for embedding extraction, taken from non-overlapping speech when available; if absent (e.g., orange), all active regions are used.} 
    \label{fig:spk_ee_strategy} 
\end{figure}

\vspace{-2mm}
\subsection{Global labels reassigmnet}
\label{sub:reassignment}
\vspace{-1mm}

After clustering, each embedding is assigned a global speaker identity. Although these could, in principle, be mapped directly to local speakers, the pipeline always performs a reassignment step, either non-constrained or constrained.  

Formally, let \( \mathbf{M}_w \in \mathbb{R}^{L \times G} \) denote the \textit{soft score matrix} for a time window \(w\), where \(L\) is the number of local speakers detected in the window and \(G\) is the number of global clusters. The entry \( M_w(l,g) \) represents the cosine similarity between the embedding of local speaker \(l\) and the centroid of cluster \(g\). 

In the non-constrained variant, each local speaker \(l\) is assigned to the global cluster \(g\) with the highest similarity score \(M_w(l,g)\) in window \(w\). This can cause multiple local speakers within the same window to collapse onto the same global identity. Alternatively, a constrained assignment based on the Hungarian algorithm~\cite{3800020109} is applied, enforcing a one-to-one mapping between active local speakers and global clusters while preserving the speaker distinctions produced by EEND. The algorithm maximizes the sum of the selected scores under the constraint that each active local speaker is assigned to a distinct global cluster. For example, when \(L=2\) local speakers are active and \(G=5\) global clusters are available,  

\vspace{-0.25cm}
\begin{equation}
\mathbf{M}_{w} =
\begin{bmatrix}
1.1 & 1.2 & \fcolorbox{blue}{white}{1.6} & \colorbox{yellow!50}{1.5} & 1.1 \\
1.0 & 1.3 & \fcolorbox{blue}{yellow!50}{1.7} & 1.0 & 1.2
\end{bmatrix},
\label{eq:constrain_example}
\end{equation}
the \colorbox{yellow!50}{constraint} ensures that the two local speakers are assigned to different global clusters, whereas the \fcolorbox{blue}{white}{non-constrained} variant may assign both to the same cluster.

As confirmed by the pyannote authors, the original implementation of the constrained variant contained an error that led to unintended behavior: embeddings corresponding to inactive speakers 
were incorrectly included in the $M_w$ matrices. To clearly separate this behavior, we denote the baseline method with ``pya-c'' prefix, while the corrected constrained version used in our proposed solution is simply denoted with ``c'' prefix (e.g. constained agglomerative hierarchical clustering (cAHC), that uses contrained variant for reassigment, same for cVBx). A detailed analysis of the behavior of ``pya-c'' is provided in Section~\ref{sec:analysis}.


\vspace{-2mm}
\subsection{Stitching and post-processing} 
\label{sub:stiching}
\vspace{-1mm}

In the stitching step, the overlapping local diarization windows illustrated in Figure~\ref{fig:from_local_to_global}~(a) are aggregated into a consistent global output, shown in Figure~\ref{fig:from_local_to_global}~(d). For each frame $f$, the overlapping windows vote on the number of active speakers, as shown in Figure~\ref{fig:from_local_to_global}~(b). Majority voting determines the estimated number of active speakers $n_f$ at frame $f$, after which the $n_f$ most frequent speakers are retained, as depicted in Figure~\ref{fig:from_local_to_global}~(c). The frame-level predictions are then merged into contiguous speech segments, producing the global diarization output in Figure~\ref{fig:from_local_to_global}~(d). 

Finally, an optional \textit{post-processing} step fills within-speaker gaps shorter than a predefined duration~$\Delta$, improving temporal consistency.


  \definecolor{highlightcf}{HTML}{FFF2CC}
\begin{figure}[t]
  \centering
  \begingroup
  \captionsetup{font=small,aboveskip=1pt,belowskip=1pt}
  \captionsetup[sub]{font=small,aboveskip=0pt,belowskip=1pt}

  \begin{subfigure}{0.99\linewidth}
    \centering
    \captionsetup{justification=raggedright,singlelinecheck=false}
    \caption{Outputs of local EEND for overlapping windows with global speaker labels indicated by colors.}
    
    \scalebox{1}[0.8]{\includegraphics[width=\linewidth]{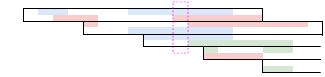}}
    \label{fig:mv1}
  \end{subfigure}

  \vspace{-12pt}

  \begin{subfigure}{0.99\linewidth}
    \centering
    \captionsetup{justification=raggedright,singlelinecheck=false}
    \caption{For each frame $f$, the row with the maximum count \colorbox{highlightcf!50}{$n_f$} is the estimated number of active speakers. White cells denote 0 counts.}
    \scalebox{1}[0.8]{\includegraphics[width=\linewidth]{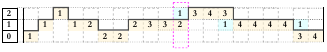}}
    \label{fig:mv2}
  \end{subfigure}

  \vspace{-10pt}

  \begin{subfigure}{0.99\linewidth}
    \centering
    \captionsetup{justification=raggedright,singlelinecheck=false}
    \caption{For each frame $f$, pick top $n_f$ speakers with the highest counts.}
    \scalebox{1}[0.8]{\includegraphics[width=\linewidth]{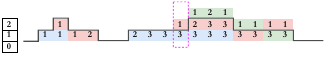}}
    \label{fig:mv3}
  \end{subfigure}

  \vspace{-10pt}

  \begin{subfigure}{0.99\linewidth}
    \centering
    \captionsetup{justification=raggedright,singlelinecheck=false}
    \caption{Global diarization output of previous step with 3 speakers.}
    \scalebox{1}[0.6]{\includegraphics[width=\linewidth]{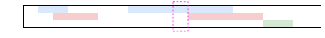}}
    \label{fig:mv4}
  \end{subfigure}

  \vspace{-5pt}
  \caption{Illustration of the stitching process that aggregates local diarization decisions into a coherent global diarization output.}
  \label{fig:from_local_to_global}
  \endgroup
\end{figure}

\section{Proposed clustering improvements}

\subsection{Bayesian clustering by GMM-VBx}
\label{sec:vbx_prop}
\vspace{-1mm}

VBx~\cite{landini2022bayesian} is a Bayesian clustering method for sequences of x-vectors, formulating diarization as probabilistic inference in an HMM with state distributions derived from a pre-trained PLDA model that captures between- and within-speaker variability. The method infers the number of speakers automatically by driving the learned prior probabilities of redundant speakers to zero. In this work, the original HMM topology is simplified by setting the transition probabilities of staying in the same speaker state ($P_{\text{loop}}$) to zero, which effectively turns the model into a Gaussian Mixture Model (GMM) \cite{klement2024discriminative}.

In the context of EEND-VC, this simplification is necessary because embeddings are not extracted from contiguous speech segments, but from portions of speech (preferably non-overlapping) assigned to each local speaker within a time window. Consequently, a single embedding can concatenate multiple disjoint waveform segments as illustrated in Figure~\ref{fig:spk_ee_strategy}, and the sequential continuity assumed by HMMs no longer holds.  

This reduction is consistent with previous cascade-based diarization systems, where GMM formulations have been shown to perform comparably to full HMMs while reducing computational cost. It also integrates naturally with the baseline AHC that uses UPGMC (centroid linkage) as initialization for the VBx algorithm. Since UPGMC produces hundreds of clusters, switching to the GMM-based variant makes VBx simpler and significantly faster \cite{klement2024discriminative}, enabling it to handle a large number of initial speakers efficiently. 

As with AHC, we can introduce the label reassignment with the constrained variant procedure explained in Section \ref{sub:reassignment}. We call this version cVBx.

\vspace{-2mm}
\subsection{Filtering}
\label{sec:filtering}
\vspace{-1mm}

We filter out unreliable embeddings extracted from very short segments before clustering, as such segments often contain overlapped speech and provide poor speaker representation. We only keep the higher-quality embeddings extracted from segments longer than a threshold, $E$. 
The remaining, higher-quality embeddings are then passed to the clustering algorithm (AHC, VBx), which produces more reliable estimates of speaker centroids. 

In the subsequent reassignment step (Section~\ref{sub:reassignment}), the similarity matrices \( \mathbf{M}_w \) are calculated with all embeddings, but since the centroids were now estimated from higher quality data, the diarization performance can be improved. 

The overall process is as follows. 
We first perform local diarization with EEND and extract speaker embeddings. Short segments are removed before an initial AHC clustering (without \(mcs\)) to initialize VBx. VBx is then applied, followed by constrained global label reassignment (Sec. \ref{sub:reassignment}) which also takes care of re-inserting short segments. Finally, stitching and post-processing are performed.
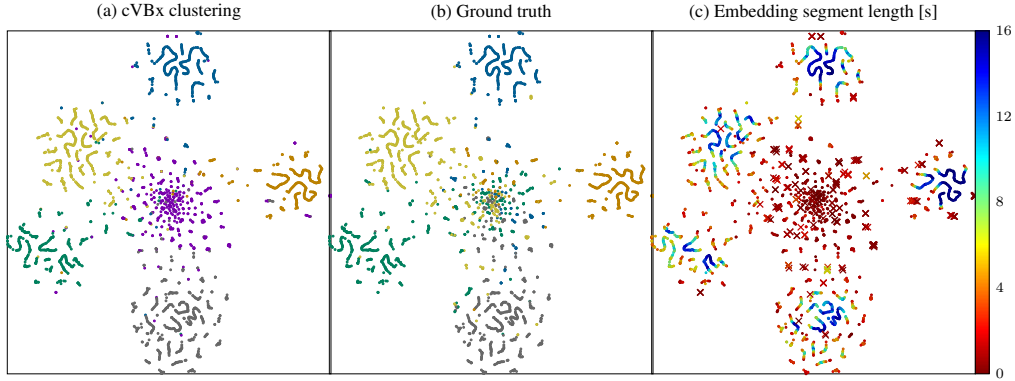
\begin{figure*}[t]
\centering
\resizebox{0.81\textwidth}{!}{\input{figures/tsne_plots}}
\vspace{-0.3cm}
\caption{Three t-SNE plots of embeddings: (a) VBx clustering of all extracted embeddings, followed by constrained reassignment, which resulted in six speakers; (b) ground truth labels, where each embedding is assigned to the speaker contributing the majority of speech in its segment (five speakers in total); and (c) segment duration in seconds indicated by color, and cross mark represent embedding from overlap.}
\label{fig:tsne}
\vspace{-0.1cm}
\end{figure*}

\section{Experimental setup}
\label{sec:experimental_setup}
\vspace{-2mm}
\subsection{Datasets}
\vspace{-1mm}

We use the training partitions of the AMI~\cite{carletta2005ami}, AISHELL-4 (AIS-4)~\cite{fu21b_interspeech}, AliMeeting (AliM)~\cite{yu2022m2met}, NOTSOFAR-1 (NSF)~\cite{vinnikov24_interspeech}, MSDWild (MSD)~\cite{liu22t_interspeech}, DIHARD3 full (DH3)~\cite{ryant21_interspeech}, RAMC~\cite{yang22h_interspeech}, and VoxConverse (VoxC)~\cite{chung20_interspeech} datasets. 
The corresponding development partitions are exclusively used for tuning the hyperparameters of the clustering methods, and then one setting of such parameters is used for evaluation on all of the test set partitions. For exact data splits, statistics, and characteristics of these datasets, see \cite{han2025efficientgeneralizablespeakerdiarization}.  

\vspace{-2mm}
\subsection{Configuration} 
\vspace{-1mm}

The EEND component in DiariZen \cite{han2025leveraging} primarily consists of WavLM \cite{9814838} and a Conformer encoder \cite{gulati20_interspeech}. The model is trained using the powerset loss \cite{plaquet23_interspeech}, with support for up to four speakers and a maximum of two overlapping speakers.
In the first set of experiments (Section~\ref{sec:analysis}), we use the model from \cite{han2025leveraging}, where the WavLM Base+ is applied and the model is trained on 8-second audio chunks from only AMI, AISHELL-4, and AliMeeting datasets.
For further evaluation (Section~\ref{sec:comparision}), we adopt WavLM Large as the encoder, where 80\% of the model parameters are removed via structured pruning \cite{louizos2018learning, han2025fine}. This variant is trained on 16-second chunks from the training partitions of all datasets. Additional details about this model can be found in \cite{han2025efficientgeneralizablespeakerdiarization}.

For inference, we adopt the configuration of \cite{han2025leveraging} and extract speaker embeddings with the ResNet34-LM from the Wespeaker toolkit \cite{wang2023wespeaker}, trained on the VoxCeleb2 dataset \cite{chung18b_interspeech}.
The PLDA model used by VBx is trained on the same VoxCeleb2 data.
System performance is evaluated using diarization error rate (DER) without collars, and we also report macro-averaged DER to capture overall performance across datasets. Since the local EEND module is fixed across experiments, only the speaker confusion component of DER can be reduced through improved clustering. A secondary reported metric is the mean speaker count error (MSCE), defined as $\mathrm{MSCE} = \tfrac{1}{R}\sum_{r=1}^{R} \lvert \hat{C}_r - C_r \rvert$, where $C_r$ and $\hat{C}_r$ denote the true and estimated number of speakers for a recording \(r\), and \(R\) is the total number of recordings in the dataset. 
Our source code is publicly available\footnote{https://github.com/BUTSpeechFIT/DiariZen}.


 \begin{table}[t]
\caption{DER (\%) on AMI, AISHELL-4, and AliMeeting datasets. Models are trained with 8s local EEND; \emph{window size}~W column indicates the inference window (8s or 16s). The last column reports average DER and MSCE in parentheses.}
\vspace{-0.3cm}
\centering
\resizebox{\linewidth}{!}{%
\begin{tabular}{c | l | l | c c c | c}
\toprule
W & Description & Method & AMI & AIS-4 & AliM & Average \\
\midrule
\multirow{4}{*}{8s}
  & Baseline             & pya-cAHC & 15.5 & 11.9 & 17.0 & 14.8 (0.83) \\
  & Corrected            & cAHC     & 17.8 & 12.3 & 17.7 & 15.9 (0.58) \\
  & Clust. $E>2s$ & cAHC     & 16.1 & 11.5 & 15.4 & 14.3 (0.16) \\
  & Clust. $E>2s$ & cVBx     & 15.7 & 11.8 & 15.2 & 14.2 (0.21) \\
\midrule
\multirow{3}{*}{16s}
  & \multirow{3}{*}{Clust. $E>4s$}
                        & cAHC     & 16.1 & 11.3 & 14.4 & 13.9 (0.20) \\
  &                     & cAHC-ASC & 15.4 & 10.8 & 14.3 & 13.5 (0.06) \\
  &                     & cVBx     & 15.5 & 10.9 & 14.2 & 13.5 (0.16) \\
\bottomrule
\end{tabular}}
\label{tab:clustering_results_merged}
\end{table}

\newcommand{\dshead}[4]{ 
  \makecell{\textbf{#1}\\ \small\color{darkgray!75}{ #3 / #4}} 
}

\newcommand{\ccolspace}{1.6cm}
\begin{table*}[t]
\centering
\caption{Column headers indicate test sets statistics with average recording length in \textcolor{darkgray!75}{minutes} and speaker range \textcolor{darkgray!75}{min–max}. The table reports DER (\%) and, for baseline and proposed methods, mean speaker count error (MSCE) in parentheses. Reported SOTA results are included for reference but may be optimistic, as they are typically obtained with dataset-specific fine-tuning.}
\label{tab:clustering_sota}
\vspace{-0.3cm}
\resizebox{0.95\textwidth}{!}{
\begin{tabular}{l | 
C{\ccolspace} C{\ccolspace} C{\ccolspace} C{\ccolspace} C{\ccolspace} C{\ccolspace} C{\ccolspace} C{\ccolspace} 
| c }
\toprule
System & 
\dshead{AMI}{16}{34.1}{3--4} &
\dshead{AIS-4}{20}{38.2}{5--7} &
\dshead{AliM}{20}{32.4}{2--4} &
\dshead{NSF}{160}{6.3}{3--7} &
\dshead{MSD}{490}{1.2}{2--4} &
\dshead{DH3}{259}{7.7}{1--9} &
\dshead{RAMC}{43}{28.8}{2--2} &
\dshead{VoxC}{232}{11.3}{1--21} &
Average \\
\midrule
pyannote v3.1 & 22.4 & 12.2 & 24.4 & / & 25.3 & 21.7 & 22.2 & 11.3 & 19.9 \\
\ \ \ + cVBx & 20.3 & 11.7 & 19.9 & / & 22.8 & 20.2 & 20.8 & 11.2 & 18.1 \\
\midrule
DiariZen Large & 15.1 (1.06) & 9.9 (1.00) & 15.5 (0.55) & 20.9 (0.59) & 18.6 (0.72) & 15.6 (0.55) & 11.1 (0.12) &  \phantom{0}9.5 (1.67) & 14.5 (0.78) \\
\ \ \ + cAHC-ASC     & 13.9 (0.15) & 9.8 (0.00) & 12.4 (0.31) & 19.4 (0.47) & 18.5 (0.67) & 15.0 (0.43) & 11.0 (0.02) & 10.1 (1.89) & 13.8 (0.49) \\
\ \ \ + cVBx      & 13.9 (0.31) & 9.9 (0.25) & 12.4 (0.15) & 17.9 (0.43) & 15.6 (0.41) & 14.6 (0.38) & 11.0 (0.09) & \phantom{0}8.8 (0.98) & \textbf{13.0} (\textbf{0.38}) \\
\midrule
\ \ \ + VBx    & 14.2 &  9.9 & 13.1 & 18.4 & 17.0 & 14.6 & 11.1 &  8.7 & 13.4 \\
\ \ \ + cVBx-nofilter & 16.6 & 10.6 & 13.9 & 18.5 & 15.5 & 15.5 & 11.2 &  9.0 & 13.9 \\
\midrule
SOTA 06/2025
& \makecell{15.4 \cite{han2025leveraging}} 
& \makecell{10.2 \cite{plaquet2025dissectingsegmentationmodelendtoend}} 
& \makecell{12.5 \cite{plaquet2025dissectingsegmentationmodelendtoend}} 
& \makecell{19.7 \cite{niu2024dcfdsdeepcascadefusion}} 
& \makecell{17.7 \cite{plaquet2025dissectingsegmentationmodelendtoend}} 
& \makecell{15.1 \cite{cheng2024sequencetosequenceneuraldiarizationautomatic}} 
& \makecell{10.7 \cite{plaquet2025dissectingsegmentationmodelendtoend}} 
& \makecell{ 9.3 \cite{plaquet2024mamba}} 
& \makecell{13.8} \\
\bottomrule
\end{tabular}
}
\end{table*}

\section{Results}
\label{sec:results}

\subsection{Analysis of proposed method}
\label{sec:analysis}
\vspace{-1mm}


Figure~\ref{fig:tsne} (a) shows the clustering with cVBx (a similar label distribution would be produced by cAHC), where an additional \emph{purple} cluster appears. The ground truth in Figure~\ref{fig:tsne} (b) does not contain such a cluster, indicating that these embeddings actually belong to some other existing speakers. The purple cluster is composed almost entirely of embeddings extracted from very short speech segments, as illustrated in Figure~\ref{fig:tsne} (c), where the color represents the duration of the concatenated waveform in seconds. Therefore, we filter these embeddings extracted from short segments before clustering and assign them afterwards, which prevents confusion and leads to a more accurate estimate of the number of speakers.

In the baseline pyannote implementation, the spurious cluster is absorbed by an inactive speaker, as a side effect of the error described in Section~\ref{sub:reassignment}. Once the one-to-one constraint is applied, the valid embeddings are redistributed to some valid speaker identities, while inactive speakers are ignored in the rest of the pipeline. 

Table~\ref{tab:clustering_results_merged} summarizes DERs on AMI, AISHELL-4, and AliMeeting. With 8\,s local EEND windows, the pyannote baseline (pya-cAHC) reaches 14.8\% DER on average. Applying the corrected constraint (cAHC) as defined in Section~\ref{sub:reassignment} degrades performance, since spurious clusters of embeddings from short segments (cf. Fig.~\ref{fig:tsne}) are not eliminated. Filtering segments shorter than 2\,s alleviates this effect, reducing DER to 14.3\% with cAHC and 14.2\% with cVBx.  

Using 16\,s windows (while the model was trained on 8\,s windows) further improves performance thanks to higher-quality embeddings and more reliable local assignments. In this setup, cAHC with a 4\,s filter yields 13.9\% DER. We also investigated an alternative stopping criterion (cAHC-ASC), which continues the clustering process until it would merge two already established speaker clusters (such that $|C| > mcs$). cAHC-ASC reduces the macro DER and MSCE to 13.5\% and 0.06, respectively. Replacing AHC with cVBx under the same conditions achieves a comparable DER of 13.5\% (0.16 MSCE). Although filtering below 4\,s can risk discarding most of the speech from less dominant speakers, it was found to be effective for these datasets.  

Overall, the results show that (i) filtering embeddings from short segments is particularly important with shorter EEND windows, and (ii) both VBx and the AHC-ASC improve over standard AHC, providing more stable speaker count estimates.



\vspace{-1mm}
\subsection{Comparison across datasets}
\label{sec:comparision}
\vspace{-1mm}

In this section, the experimental methods described in Section~\ref{sec:analysis} are evaluated with WavLM Large and the Conformer-based DiariZen-Large model, using 16-second local EEND windows. All methods prefixed with “+” in Table~\ref{tab:clustering_sota} apply filtering with a threshold $E$ of 1.6 sec, 
unless explicitly labeled with nofilter. The absence of the prefix 'c' denotes a reassignment without enforcing the constraint described in Section \ref{sub:reassignment}.  

Table~\ref{tab:clustering_sota} first reports results for pyannote v3.1, the open-source diarization system from which the DiariZen inference pipeline is derived. Pyannote v3.1 employs BLSTM layers as the decoder on top of SincNet features. Importantly, it is not fine-tuned to any specific domain.
DiariZen Large adopts the same baseline clustering code (pya-cAHC), but replaces the local EEND module of pyannote v3.1 with our Conformer-based variant.



Next, we evaluate the proposed VBx-based clustering. The comparison between cVBx and VBx highlights the benefit of constrained reassignment.  
Contrast between cVBx and cVBx-nofilter illustrates the impact of filtering unreliable embeddings. These effects are most pronounced on datasets with longer recordings and higher overlap, such as AMI and AliMeeting. In these corpora, many recordings often involve four active local speakers in the local diarization, which complicates extraction of clean single-speaker embeddings and yields a higher proportion of very short or overlapping segments. Hence, filtering these embeddings is important.

As shown in the table, cVBx achieves a performance comparable to cAHC-ASC on long-meeting datasets (AMI, AISHELL-4, AliMeeting), while providing substantial gains on MSDWild, NOTSOFAR-1, and VoxConverse. Overall, cVBx reduces the macro-averaged DER by 1.5 points compared to the baseline. 
In VoxConverse (VoxC), short pauses are annotated as speech rather than silence. To address this, we apply the simple post-processing step described at the end of Section~\ref{sub:stiching} with $\Delta = 0.5s$, yielding an improvement of about 0.5 DER. Since this step does not modify speaker confusion, it benefits all methods equally and reflects the labeling conventions of this dataset.
Finally, the last line of Table~\ref{tab:clustering_sota} reports the state-of-the-art (SOTA) results as of June 2025. By replacing the baseline pyannote clustering with cVBx, DiariZen achieves a new state-of-the-art performance across most of the evaluated datasets.

\vspace{0.1cm}
\section{Conclusion}
\label{sec:conclusion}
We have proposed improvements to the clustering stage of the EEND-VC framework, focusing on filtering unreliable short-segment embeddings and integrating a simplified VBx approach. Our analysis showed that filtering prevents spurious clusters and stabilizes speaker count estimation, particularly with short EEND windows, while VBx followed by constrained speaker embedding reassignment (cVBx) provides robust clustering across diverse conditions. Extensive experiments on eight public benchmarks demonstrated consistent gains, with cVBx yielding notable improvements on challenging datasets such as MSDWild, NOTSOFAR-1, and VoxConverse. When combined with DiariZen-Large EEND model, our system surpassed the most recent state-of-the-art in diarization without dataset-specific fine-tuning or parameter tuning, underscoring its robustness and generalizability. Future work will explore a tighter integration of clustering with local EEND models. %
\vfill\pagebreak

\bibliographystyle{IEEEbib}
\bibliography{strings,refs}

\end{document}

%% file: figures/spk_ee_strategy.tex
\begin{tikzpicture}[xscale=0.33, yscale=0.33]

\definecolor{speaker1}{RGB}{0,128,0}    
\definecolor{speaker2}{RGB}{255,140,0}  
\definecolor{speaker3}{RGB}{200,0,0}    
\definecolor{speaker4}{RGB}{128,0,128}  

\node at (12, 8) {Chunk of input recording waveform.};
\node at (12, -0.75) {Local diarization binarized output.}; 

\draw[thick, gray] (0,0) rectangle (24,4);
\foreach \y in {1,2,3}{
    \draw[gray] (0,\y) -- (24,\y);
}

\node[anchor=south west, inner sep=0, yscale=0.5] at (-1,4.5)  {\includegraphics[width=8.7cm] {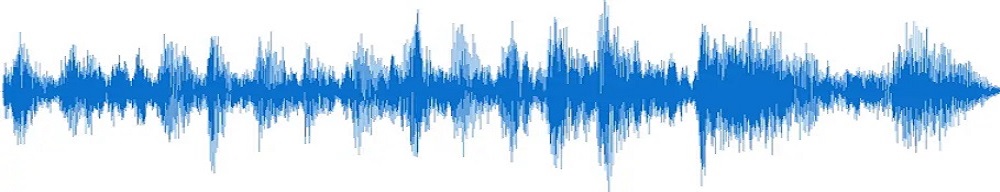}}; 

\draw[thick, speaker1] (0,4.5) rectangle (3,7.5);
\draw[thick, speaker1] (10,4.5) rectangle (12,7.5);
\draw[thick, speaker1] (18,4.5) rectangle (20,7.5);
\draw[thick, speaker2] (12.05,4.5) rectangle (17.95,7.5);
\draw[thick, speaker2] (22,4.5) rectangle (24,7.5);
\draw[thick, speaker4] (4,4.5) rectangle (8,7.5);
\draw[thick, speaker4] (21,4.5) rectangle (22,7.5);

\fill[fill=speaker1!20] (0,3) rectangle (3,4);
\fill[fill=speaker1!20] (8,3) rectangle (20,4);
\draw[thick, speaker1] (0,3) rectangle (3,4);
\draw[thick, speaker1] (10,3) rectangle (12,4);
\draw[thick, speaker1] (18,3) rectangle (20,4);
\node at (1.5,3.5) {1};
\node at (11,3.5) {2};
\node at (19,3.5) {3};
\fill[fill=speaker2!20] (12,2) rectangle (18,3);
\fill[fill=speaker2!20] (22,2) rectangle (24,3);
\draw[thick, speaker2] (12,2) rectangle (18,3);
\draw[thick, speaker2] (22,2) rectangle (24,3);
\node at (15,2.5) {1};
\node at (23,2.5) {2};
\fill[fill=speaker4!20] (4,0) rectangle (10,1);
\fill[fill=speaker4!20] (21,0) rectangle (24,1);
\draw[thick, speaker4] (4,0) rectangle (8,1);
\draw[thick, speaker4] (21,0) rectangle (22,1);
\node at (6,0.5) {1};
\node at (21.5,0.5) {2};
\end{tikzpicture}

%% file: figures/tsne_plots.tex


\newlength{\panelw}\setlength{\panelw}{\dimexpr\textwidth/3\relax}
\newlength{\panelh}\setlength{\panelh}{\panelw}

\newlength{\cbarw}\setlength{\cbarw}{0.24cm}
\newlength{\cbarsep}\setlength{\cbarsep}{6pt}
\newlength{\panelwfull}\setlength{\panelwfull}{\dimexpr\panelw+\cbarw+\cbarsep\relax}

\newlength{\threegap}\setlength{\threegap}{\dimexpr\textwidth - (2\panelw + \panelwfull)\relax}

\pgfplotsset{
  colormap={labels}{
    rgb255=(230, 159, 0)   
    rgb255=(0, 158, 115)   
    rgb255=(0, 114, 178)   
    rgb255=(240, 228, 66)  
    rgb255=(128, 128, 128) 
    rgb255=(148, 0, 211)   
  },
  every axis/.append style={
    xmin=-56.4, xmax=64.5,
    ymin=-64.3, ymax=63.9,
  },
}

\newlength{\cbarlabelw}
\settowidth{\cbarlabelw}{\scriptsize 16}
\addtolength{\cbarlabelw}{1pt} 

\setlength{\panelwfull}{\dimexpr\panelw+\cbarw+\cbarsep+\cbarlabelw\relax}
\setlength{\threegap}{\dimexpr\textwidth - (2\panelw + \panelwfull)\relax}

\noindent\makebox[\textwidth][c]{%
\begin{minipage}[t]{\panelw}
\centering
\tikzsetnextfilename{plot-I}
\begin{tikzpicture}[baseline=(axI.north)]
\begin{axis}[
  name=axI,
  width=\panelw, height=\panelh,
  scale only axis, axis equal image,
  xtick=\empty, ytick=\empty, tick style={draw=none},
  title style={at={(0.5,1.0)}, anchor=south, yshift=-1ex},
  title={(a) cVBx clustering}, 
  colormap name=labels, colormap access=piecewise constant,
  point meta min=0, point meta max=6
]
\addplot[scatter, only marks, scatter src=explicit, mark size=0.45pt]
  table[col sep=comma, x=x, y=y, meta=v]{tsne.csv}; 
\end{axis}
\end{tikzpicture}
\end{minipage}\hspace{0.5\threegap}%

\begin{minipage}[t]{\panelw}
\centering
\tikzsetnextfilename{plot-II}
\begin{tikzpicture}[baseline=(axII.north)]
\begin{axis}[
  name=axII,
  width=\panelw, height=\panelh,
  scale only axis, axis equal image,
  xtick=\empty, ytick=\empty, tick style={draw=none},
  title style={at={(0.5,1.0)}, anchor=south, yshift=-1ex},
  title={(b) Ground truth}, 
  colormap name=labels, colormap access=piecewise constant,
  point meta min=0, point meta max=6
]
\addplot[scatter, only marks, scatter src=explicit, mark size=0.45pt]
  table[col sep=comma, x=x, y=y, meta=g]{tsne.csv}; 
\end{axis}
\end{tikzpicture}
\end{minipage}\hspace{0.5\threegap}%

\begin{minipage}[t]{\panelwfull}
\centering
\tikzsetnextfilename{plot-III}
\begin{tikzpicture}[baseline=(axIII.north)]
\begin{axis}[
  name=axIII,
  width=\panelw, height=\panelh, 
  scale only axis, axis equal image,
  xtick=\empty, ytick=\empty, tick style={draw=none},
  title style={at={(0.5,1.0)}, anchor=south, yshift=-1ex},
  title={(c) Embedding segment length [s]}, 
  colormap name=myrevjet,
  point meta min=0, point meta max=16,
  colorbar,
  colorbar style={
    width=\cbarw,
    height=\pgfkeysvalueof{/pgfplots/parent axis height},
    at={(1,0.5)}, anchor=west,
    ytick={0,4,8,12,16},
    yticklabel style={font=\scriptsize, text height=1.2ex, text depth=.25ex}, 
  },
   outer sep=0pt, 
]

\addplot[
  scatter, only marks, scatter src=explicit,
  mark=*, mark size=0.45pt,
  restrict expr to domain={\thisrow{o}}{0:0},
  unbounded coords=discard, filter discard warning=false,
] table[col sep=comma, x=x, y=y, meta=l]{tsne.csv};

\addplot[
  scatter, only marks, scatter src=explicit,
  mark=x, mark size=2.2pt, line width=0.6pt,
  restrict expr to domain={\thisrow{o}}{1:1},
  unbounded coords=discard, filter discard warning=false,
] table[col sep=comma, x=x, y=y, meta=l]{tsne.csv};


\end{axis}
\end{tikzpicture}
\end{minipage}%
}